\documentclass[traditabstract]{aa}  
\usepackage[varg]{txfonts}
\usepackage{graphicx}
\usepackage{subfigure}
\usepackage{rotating}
\usepackage{color}
\usepackage{natbib}
\usepackage{float}
\usepackage{wrapfig,lipsum}
\bibpunct{(}{)}{;}{a}{}{,} 
\definecolor{darkgreen}{rgb}{0.0,0.5,0.0}
\definecolor{darkred}{rgb}{0.5,0.0,0.0}
\definecolor{grey}{rgb}{0.4,0.5,0.6}

\begin{document} 
\title{Ultimate Merging at z$\sim$0.1}
\titlerunning{Ultimate Merging at z$\sim$0.1}
\authorrunning{Maschmann et al.}

\author{Daniel Maschmann\inst{1,2}, Anne-Laure Melchior\inst{1}}
\institute{\,Sorbonne {Universit\'e}, LERMA, Observatoire de Paris, PSL research university, CNRS, F-75014, Paris, France
\and 
\,RWTH Aachen University, Institute for Philosophy of Science and Technology, Aachen, Germany\\
\email{Daniel.Maschmann@rwth-aachen.de, A.L.Melchior@obspm.fr}
}
\date{Received April ?, 2019; accepted ?}
\abstract {We present a study of 58 double-peaked emission line galaxies for which  one of the components is suppressed in ${\rm [OIII]}\lambda$5008 or significantly weaker than the other one. Accordingly, the two components are classified differently in the BPT diagram. We show that the strong ${\rm [OIII]}$  component coincides with the stellar velocity and the suppressed component is off-centred in 66$\%$ of the galaxies, while in 12$\%$ of them it is the opposite.
The analysis of their morphology reveals that about half of the sample is composed of S0, the rest is composed in equal part of mergers and late-type galaxies. We discuss that these characteristics exclude rotating discs and suggest different stages of merging. It is possible that the number of mergers is underestimated if the double nuclei are not resolved. Tidal features are detected in the outskirts of some S0 galaxies. This high fraction of S0 is surprising, as in addition most of the galaxies are isolated and the others in small groups. All these galaxies, hosting an AGN component, are massive, lie on the star forming sequence, and exhibit an enhanced star formation in their centre. While we cannot exclude outflows, these galaxies exhibit a  spectra, which do not correspond to usual outflow observations characterised by high gas velocities, and the standard deviations of the two peaks are comparable. In parallel, these characteristics are compatible with ultimate stages of galaxy merging, where the two nuclei are too close to be detected or dynamical disturbances might be present in post-mergers like massive S0.}
\maketitle
\section{Introduction}
The building-up of galaxies is currently understood as a co-evolution of the star formation (SF), usually strong in the central parts, and the accretion activity of the central black hole  \citep{2014ARA&A..52..415M}. While galaxy interactions can trigger SF \citep{1994ApJ...425L..13M,1996ApJ...464..641M,2013A&A...557A..59B}, gas accretion is not negligible \citep{2005MNRAS.363....2K,2008A&ARv..15..189S}. In the local Universe, interactions and minor/major mergers do not necessarily enhance SF (e.g. \citet{2003A&A...405...31B,2005ApJ...625...23B,2007A&A...468...61D}), but they can also drive gas towards the centre, fuel the nuclear black hole, enhancing active galactic nuclei (AGN) activity and feedback \citep{2006MNRAS.365...11C,2005MNRAS.361..776S}. Major mergers are definitely providing the largest starbursts \citep{1996ARA&A..34..749S}, but they are rare, and their influence on the total cosmic SF is low \citep{2007A&A...468...61D,2009ApJ...704..324R}.
Galaxy mergers are usually identified on their optical morphology. Wide field observations connect enhanced AGN and starburst activities with interactions in galaxy pairs \citep{2013MNRAS.433L..59P,2014MNRAS.441.1297S}.
\begin{figure}
  \centering 
 \includegraphics[width=0.48\textwidth]{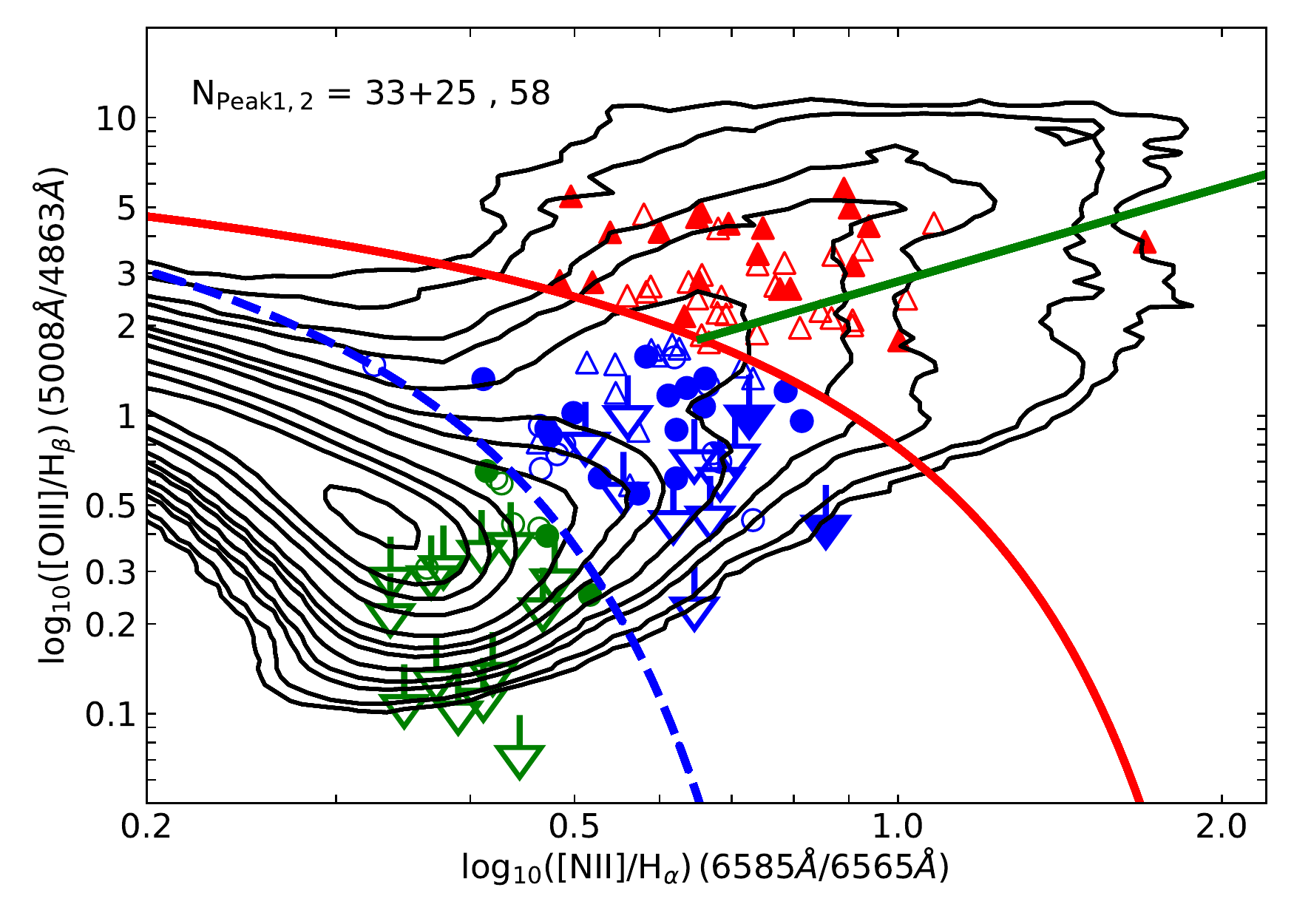}
  \caption{BPT diagnostic diagram \citep{2006MNRAS.372..961K} with topological classification based on  \citet{2003MNRAS.346.1055K}, \citet{2001ApJ...556..121K} and \citet{2007MNRAS.382.1415S}. We show the two components of the double-peak galaxies: 58 triangles represent the strong {\rm [OIII]} component and 33 circles the weaker {\rm [OIII]} component. We display the 25 upper limits with an arrow for those galaxies, which have a SNR$<$3 suppressed component. We highlight galaxies with a off-centred weak {\rm [OIII]} line with an empty marker (see Sect. \ref{ssec:kinematics}).}
  \label{fig:first:bpt}
\end{figure}
\begin{figure*}
    \includegraphics[width=0.95\linewidth]{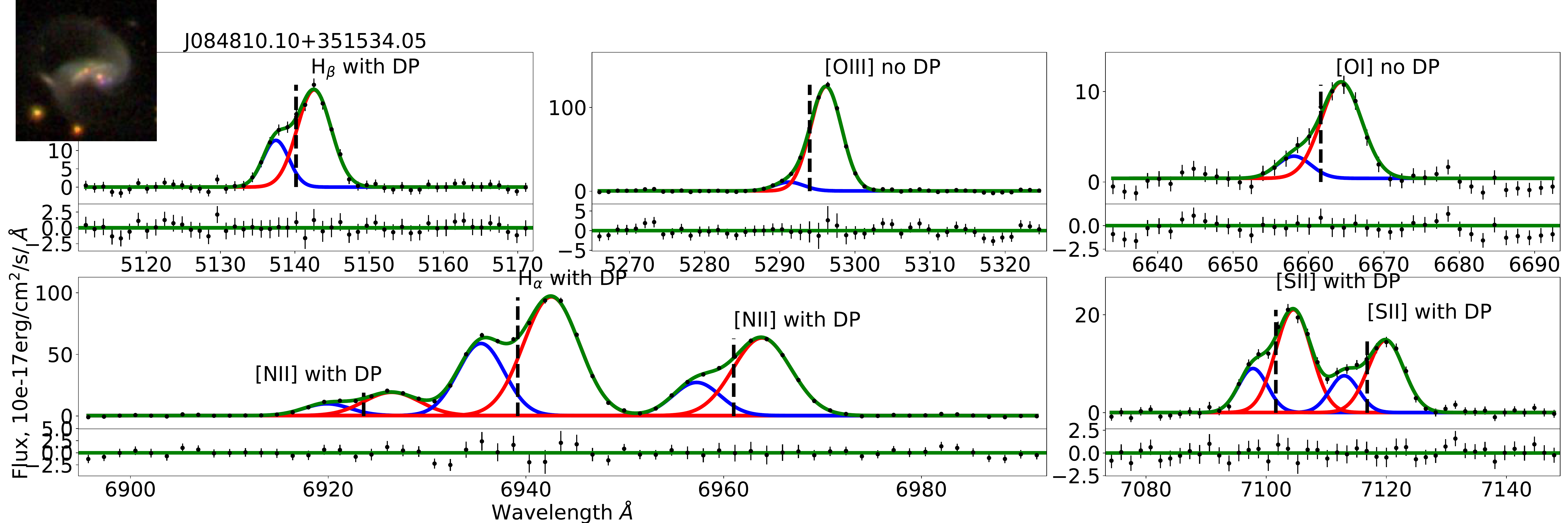}
    \caption{Emission lines of one double-peak galaxy classified as merger \citep{2018MNRAS.476.3661D} at z=0.057, namely ${\rm H}_{\beta}\lambda$4863, {\rm [OIII]}$\lambda$5008, $\rm [OI]\lambda$6302, ${\rm [NII]}\lambda$ 6550, ${\rm H}_{\alpha}\lambda$6565, ${\rm [NII]}\lambda$ 6586, ${\rm [SII]}\lambda$6718 and ${\rm [SII]}\lambda$ 6733. We display on the top left, the SDSS snap-shot. Each displayed line is fitted with a double Gaussian function (with velocities fixed with the stacked spectra) as displayed by blue and red lines. The black dashed line indicates the position of the stellar velocity of the host galaxy, computed by \citet{2017ApJS..228...14C}. As in Maschmann et al. (in prep), we indicate emission lines with a confirmed double-peak "with DP" (resp. "no DP"). Beside SNR constraints, the non-detection corresponds to a line weaker than a factor 3.}
    \label{fig:Spec}
\end{figure*}
\begin{figure*}
  \centering 
 \includegraphics[width=0.95\textwidth]{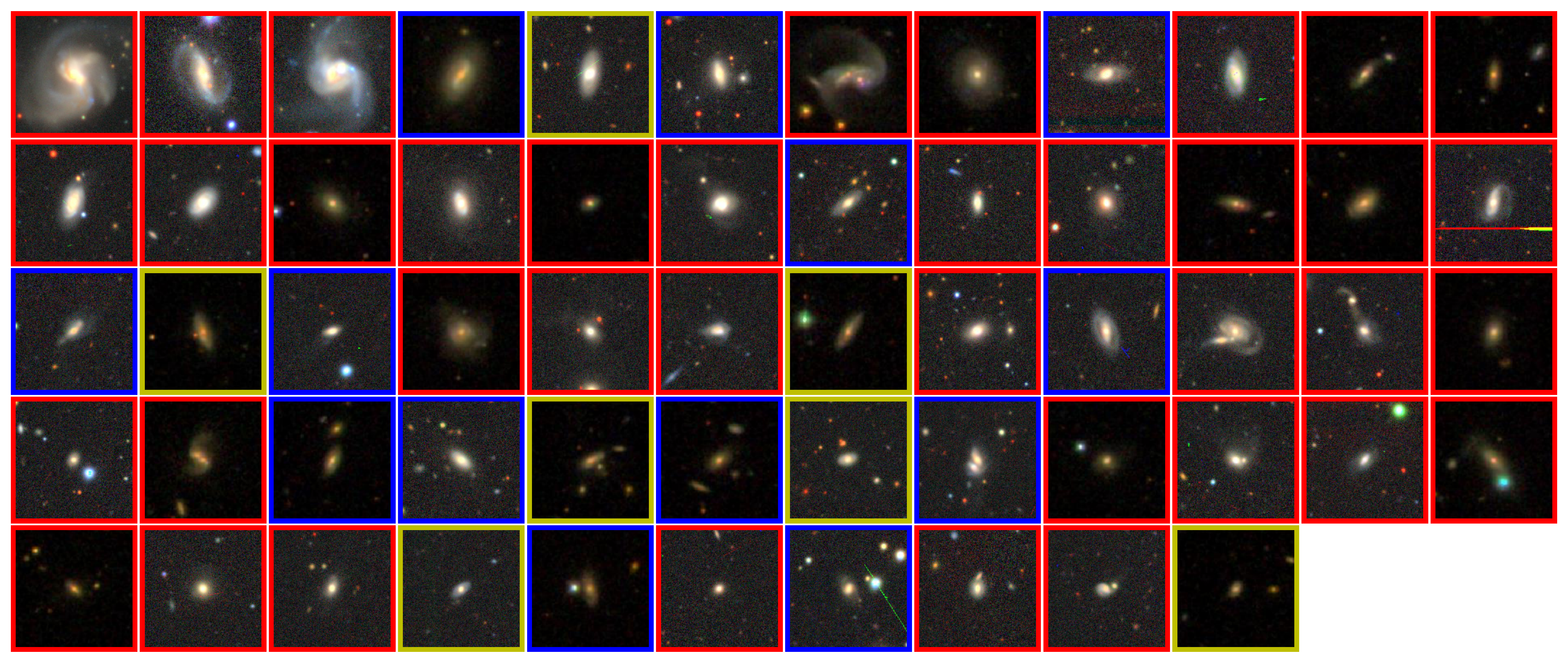}
  \caption{Galaxy $150^{\prime\prime}\times 150^{\prime\prime}$ snapshots sorted by redshift. We show the Legacy Survey snapshots \citep{2019AJ....157..168D} if available otherwise we display the SDSS snapshots \citep{2002AJ....124.1810S}. We highlight galaxies which are classified with a centred strong (resp. weak) {\rm [OIII]} component on the stellar velocity with a red (resp. yellow) and those with symmetrical kinematics with blue frames as described in Sect. \ref{ssec:kinematics}.}
  \label{fig:images}
\end{figure*}

In this letter, we investigate a galaxy sample based on a spectroscopic selection. Relying on the RCSED catalogue from \citet{2017ApJS..228...14C}, Maschmann et al. (in prep.) have performed an automatic selection of about 5000 galaxies from the SDSS catalogue \citep{2002AJ....124.1810S}, exhibiting double-peak emission lines. We study here 58 such double-peak galaxies, whose peaks exhibit two different Baldwin, Phillips $\&$ Terlevich (BPT) diagnostics: AGN/SF, AGN/composite or composite/SF. Previous works on double-peaked emission-lines, motivated by dual or offset AGNs, were restricted to AGN lines. Different works identified galaxy mergers and AGN outflows \citep{2018ApJ...867...66C,2018ApJ...854..169L,2018MNRAS.473.2160N,2015ApJ...813..103M}. As discussed in \citet{2012ApJS..201...31G}, beside dual AGN, double-peaked emission lines can correspond to several configurations difficult to disentangle: rotating disk, gas outflow or different gas components due to a galaxy merger. In this work, we focus on double-peak emission-line galaxies with a high signal-to-noise ratio (SNR) in all lines, but with one of the {\rm [OIII]}$\lambda$5008 components absent or significantly weaker than the other one. As further argued throughout this letter, we do not expect rotating discs exhibiting such a pattern, which favours the merging and possibly outflow scenarios. In Sect. \ref{sect:data}, we present the spectroscopic selection of the galaxy sample, their host properties, the estimated star formation and their environment. In Sect.  \ref{sect:disc}, we discuss the results. 
A cosmology of $\Omega_{m} = 0.3$, $\Omega_{\Lambda} = 0.7$ and $h = 0.7$ is assumed in this work.

\section{Data analysis: host properties of merger candidates}
\label{sect:data}
Relying on the RCSED catalogue \citep{2017ApJS..228...14C}, a single and a double Gaussian function are fitted to the emission-lines and several selection criteria have been applied. We stacked all fitted emission lines and introduced a global peak position and velocity dispersion ($\sigma$) for each peak. $\phi_1$ (resp. $\phi_2$) corresponds to the flux of the blueshifted (resp. redshifted) component. The catalogue finally provides parameters of the fitted double Gaussian function for the different optical lines. We restrict the analysis to the ${\rm H}_{\beta}\lambda$4863, {\rm [OIII]}$\lambda$5008, ${\rm H}_{\alpha}\lambda$6565 and ${\rm [NII]}\lambda$ 6586 emission-lines.

\subsection{Spectroscopic selection and classification}
\label{ssct:spectroscopic:selection}
We focus here on galaxies with (1) a ${\rm H}_{\alpha}\lambda$6565 flux ratio between the two components  in the range $1/2 < \phi_1 /\phi_2 < 2$ and (2) an {\rm [OIII]}$\lambda$5008 flux ratio below $1/3$ or above $3$. We also require a signal-to-noise ratio (SNR) larger than 10 for the {\rm [OIII]}$\lambda$5008 line. We hence select galaxies with a well-defined double-peak structure in the ${\rm H}_{\alpha}\lambda$6565 line, but suppressed in {\rm [OIII]}$\lambda$5008.  At this stage, we select 123 galaxies. We then classify each emission line component individually with a BPT diagram \citep{1981PASP...93....5B,2006MNRAS.372..961K} and select the galaxies with two different classifications. We finally get 58 galaxies as displayed in Figure\,\ref{fig:first:bpt}. The weaker components are situated in the composite and star forming regions. The components with the higher {\rm [OIII]} flux are clearly associated with AGN/LINER activity but for 11 located in the composite region. We later refer to the two components as the weak {\rm [OIII]} and the strong {\rm [OIII]} components. Examples of spectra of galaxies thus selected are displayed in Figure\,\ref{fig:Spec} and Figure\,\ref{fig:Spec2}. 

We also classify the sample with the WHAN diagram shown in Figure\,\ref{fig:whan}. While a shift is observed between the two components, there are only 9 out of 58 galaxies with AGN/SF classification. The BPT-AGN are classified strong and weak AGN in the WHAN diagnostics. The BPT-based composite and star forming classifications are more ambiguous.   However, the WHAN diagram \citep{2010MNRAS.403.1036C} based on equivalent widths is biased if there is an AGN in one component as the continuum of both components will be affected. 

In Table\,\ref{table:properties}, we list the properties of the 58 galaxies sorted by redshift.

\subsection{Morphology}
\label{ssec:morphology}
Our selected sample is composed of galaxies with redshifts in the range 0.04-0.17, corresponding to a SDSS 3$"$ fiber diameter between 2 and 10 kpc. The mean stellar masses are around log$_{10}$(M/M$_{\odot}$)$\sim$11 \citep{2003MNRAS.346.1055K}. Relying on a machine-learning-based morphological classification \citep{2018MNRAS.476.3661D} and a visual correction when required, we classify this sample as 15 (26$\%$) merger, 16 (28$\%$) late-type (LTG), 1 (2$\%$) early type (ETG) and 26 (45$\%$) S0 galaxies. Only one galaxy has been (mis-)classified as elliptical, and it might be a S0. Several galaxies were mis-classified by machine-learning-based morphological classification, e.g. close double nuclei were missed.  We show snapshots in Figure\,\ref{fig:images}. Beside galaxies classified mergers, many tidal features can be observed even around S0 galaxies. This can be compared to the work of \citet{2018A&A...617A.113E}, who show that S0 galaxies resulting from major/minor mergers exhibit tidal features in their outskirts. Last, one can note the large fraction of S0 galaxies compared to the usual fraction observed in magnitude limited samples e.g. $11\%$ in Shapley-Ames Catalog \citep{2009ApJ...702.1502V}.

\subsection{Kinematics}
\label{ssec:kinematics}
The velocity differences between the two peaks are between 215 and 415\,km\,s$^{-1}$. These values correspond to the upper range of the Tully-Fisher relation expected according to the stellar masses \citep{2000ApJ...533L..99M}. 

\begin{figure}
  \centering 
 \includegraphics[width=0.45\textwidth]{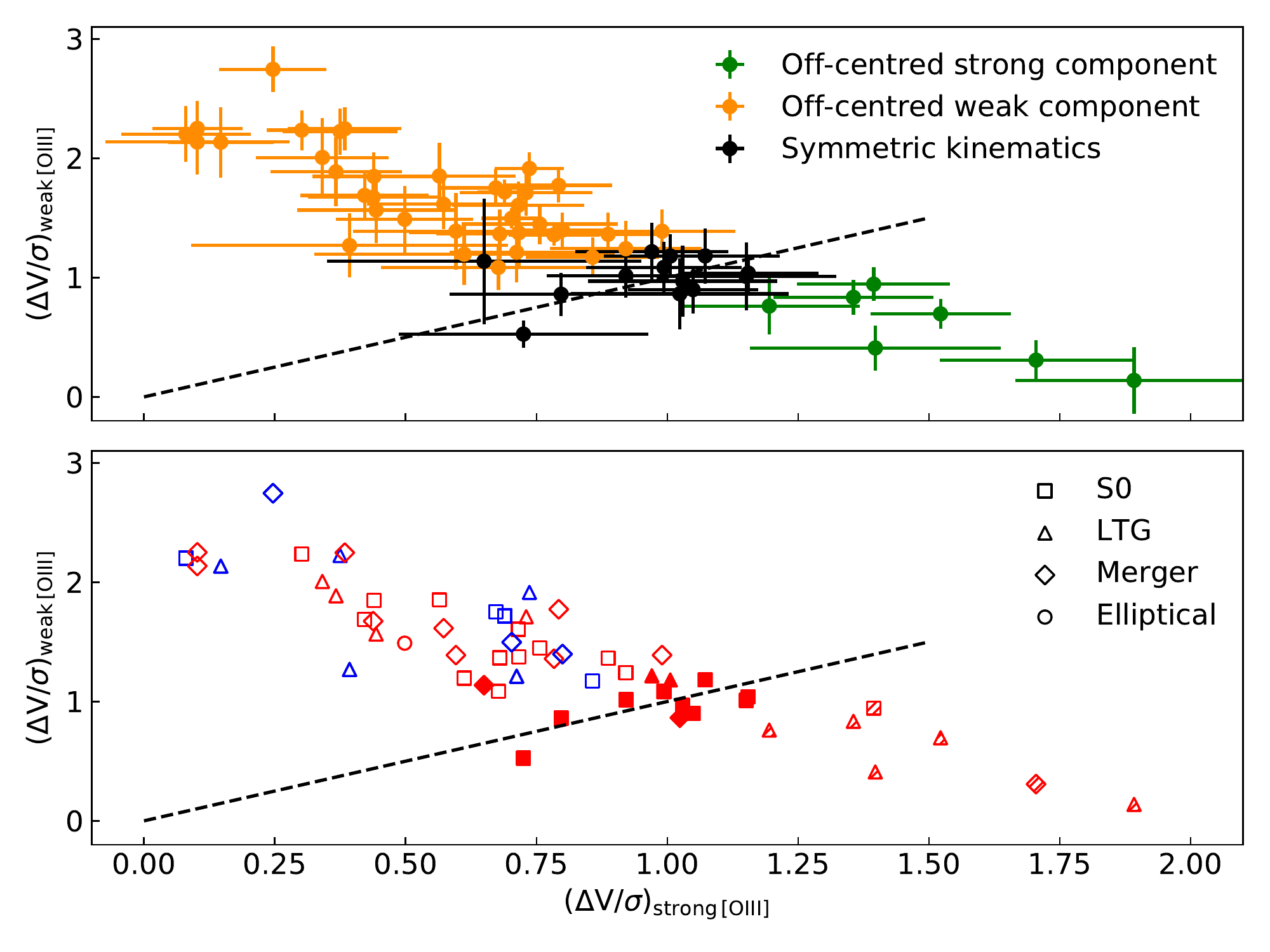}
  \caption{Velocity offsets of the two emission-line components relative to the stellar velocity of the host galaxy in units of their velocity dispersion $\sigma$. The two panels show the exact same points. On the y-axis (resp. x-axis), we display the relative offset of the weak {\rm [OIII]} (strong {\rm [OIII]}) component. In the upper panel, we compute error-bars and colour-code off-centred strong (resp. weak) {\rm [OIII]} lines in green (resp. orange) and symmetric kinematics in black. In the lower panel, we encode the morphological classification with different marker styles and the BPT classification of the strong {\rm [OIII]} component (see Figure\,\ref{fig:first:bpt}) with the colour.}
  \label{fig:offset:classification}
\end{figure}

We then compare the velocity of each emission-line component with the stellar velocity. 
We compute the ratio of  $\Delta {\rm V} = {\rm v}_{\rm peak}  - {\rm v}_{*}$ the difference between the individual peak position ${\rm v}_{\rm peak}$ and the stellar velocity ${\rm v}_{*}$ of the host galaxy, to the velocity dispersion $\sigma$ of the component. In Figure\,\ref{fig:offset:classification}, we display these ratios for the weak {\rm [OIII]} component as a function of those of the strong {\rm [OIII]} component. We then study if one of the two peaks is centred on the same velocity as the stars or not. We classify galaxies with off-centred strong (resp. weak) {\rm [OIII]} components as those showing a velocity offset in this component, larger than at least $1\,\sigma$ from the stellar velocity. We clearly see that for the majority (66\,$\%$) of these galaxies, the position of the strong {\rm [OIII]} peak is closer to the stellar velocity than for the weak {\rm [OIII]} peak. This comprises also all galaxies with a strong {\rm [OIII]} component classified as composite (see Sect. \ref{ssct:spectroscopic:selection}). But we find also 7 galaxies showing an off-centred strong {\rm [OIII]} component. The latter are mostly classified as LTG (5), and the remaining are respectively one merger and one S0 galaxy. Last, we also find that 13 galaxies ($22\%$) have the stellar velocity centred between the two peaks. The majority (9) of those are S0 galaxies, the others are 2 LTG and 2 mergers.

\subsection{Environment}
\label{ssec:env}
Following \citet{2007ApJ...671..153Y}, we estimate the number of identified neighbours, associated to each galaxy of the studied sample. The majority of the galaxies (67\,$\%$) is composed of isolated galaxies with no identified counterparts, 46\,$\%$ of those are classified S0, 21\,$\%$ mergers and 31\,$\%$ late-type galaxies. 12 galaxies are located in small groups composed of 2 to 5 galaxies. 3 (25\,$\%$) are late-type, 5 (41\,$\%$) mergers and 4 (33\,$\%$) S0 galaxies. Only one S0 galaxy is located in a small cluster with 27 counterparts. Last, 6 galaxies (10\,$\%$) have not been processed by \citet{2007ApJ...671..153Y}: 2 mergers, 3 S0 galaxies and 1 LTG. 

23 (61\,$\%$) of the galaxies with an off-centred weak {\rm [OIII]} component are isolated, while 11 (29\,$\%$) of those have counterparts (one of these is in the small cluster above mentioned), 4 (11\,$\%$) have not been processed by \citet{2007ApJ...671..153Y}. 
Regarding galaxies with an off-centred strong {\rm [OIII]} component, 4 (57\,$\%$) of them are in isolated galaxies, and 1 (14\,$\%$) in a pair of galaxies. The 2 (29\,$\%$) remaining are not in \citet{2007ApJ...671..153Y}. Last, the majority 12 (92\,$\%$) of galaxies with symmetrical kinematics are isolated and only 1 (8\,$\%$) is in a pair.

\subsection{Star formation}
\label{ssec:sf}
\begin{figure}
  \centering 
 \includegraphics[width=0.45\textwidth]{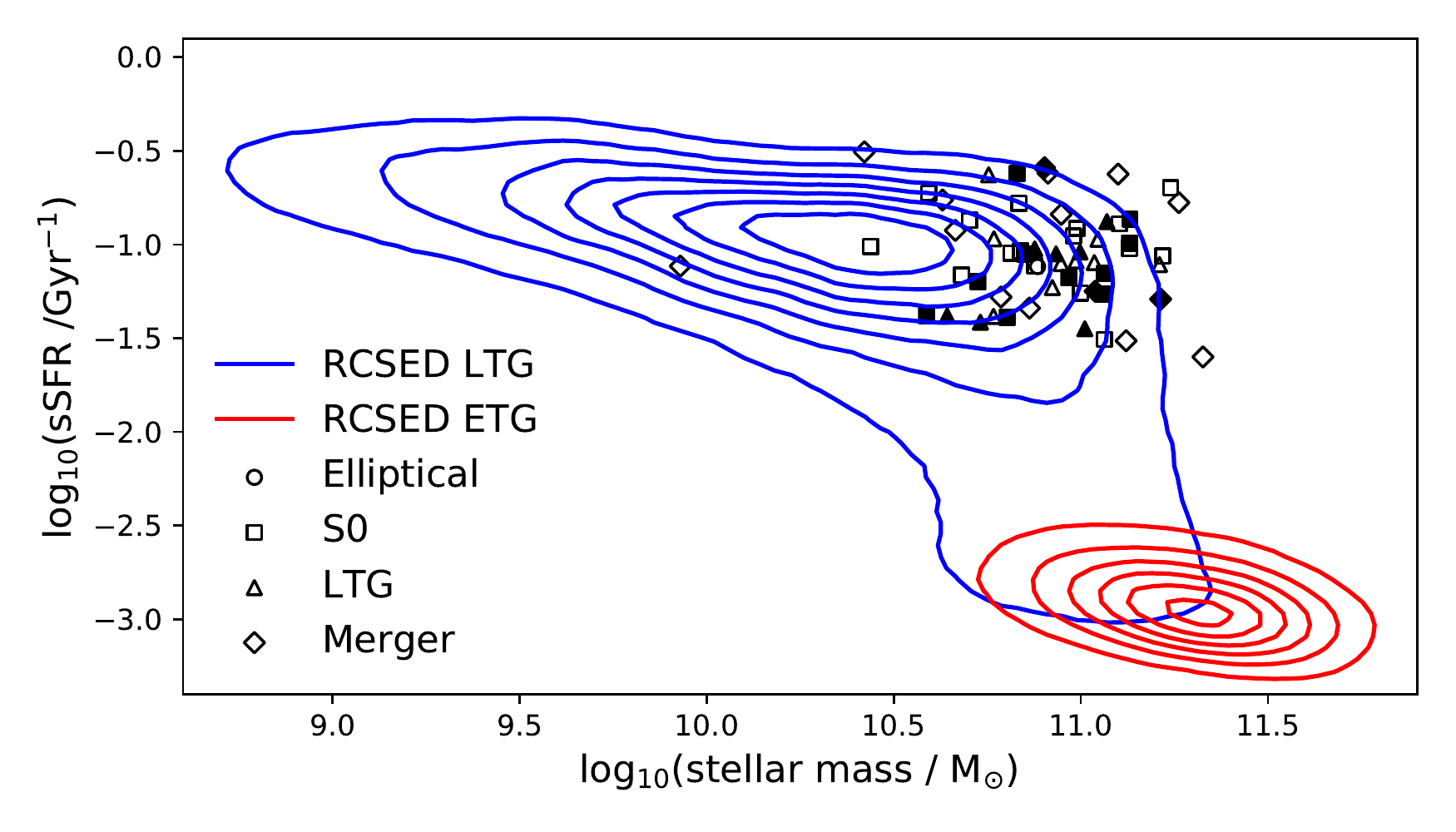}
  \caption{Specific star formation rate (sSFR) \citep{2004MNRAS.351.1151B} as a function of stellar mass \citep{2003MNRAS.346.1055K}.  We compute as blue (resp. red) contour lines a sub-sample of LTG (resp. ETG) selected from the RCSED catalogue \citep{2017ApJS..228...14C}. We display our galaxy sample with their morphology encoded in the marker style and highlight galaxies with off-centred weak {\rm [OIII]} lines by empty marker.}
  \label{fig:ssfr:stellar:mass}
\end{figure}

\begin{figure}
  \centering 
 \includegraphics[width=0.45\textwidth]{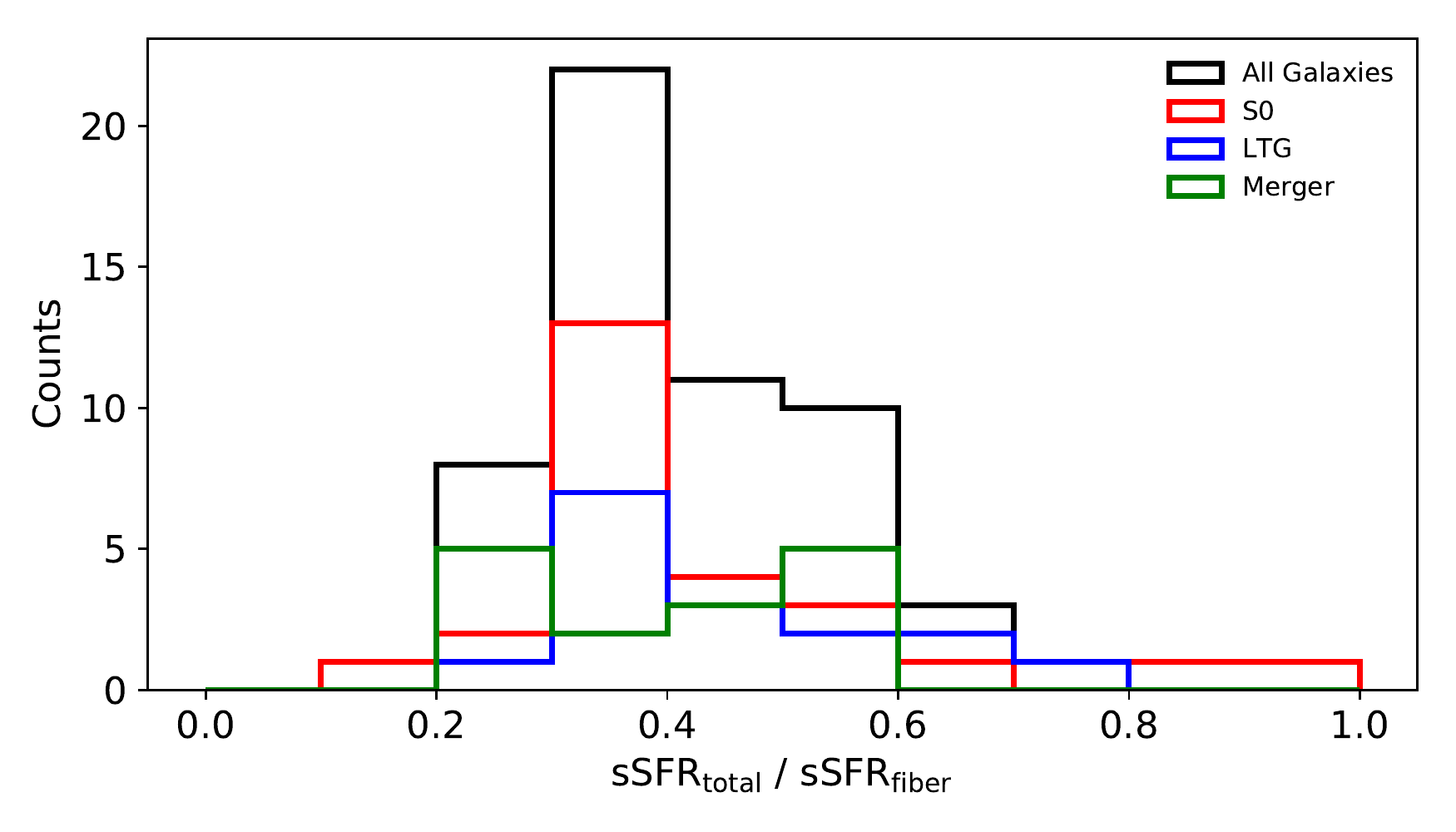}
  \caption{Ratio of the total sSFR and the sSFR measured within the 3$"$ fibre by \citet{2004MNRAS.351.1151B}. The histogram is decomposed into different morphological types (\ref{ssec:morphology}).}
  \label{fig:ssfr:ratio}
\end{figure}

In Figure\,\ref{fig:ssfr:stellar:mass}, we compute the stellar mass - sSFR diagram as discussed by \citet{2004MNRAS.351.1151B} with stellar mass (resp. sSFR) computed by \citep{2003MNRAS.346.1055K} (resp. \citet{2004MNRAS.351.1151B}).
The galaxies from our sample are not quenched and exhibit a star formation ratio typical of the upper-mass range of the main sequence. In Figure\,\ref{fig:ssfr:ratio}, we present the ratio between the specific star formation ratio (sSFR) of the total galaxy and the sSFR of the 3$"$ SDSS fiber. The sSFR measurements are computed from \citet{2004MNRAS.351.1151B}. For all the galaxies, we observe a star formation stronger in the central regions.

\begin{figure}
  \centering 
 \includegraphics[width=0.48\textwidth]{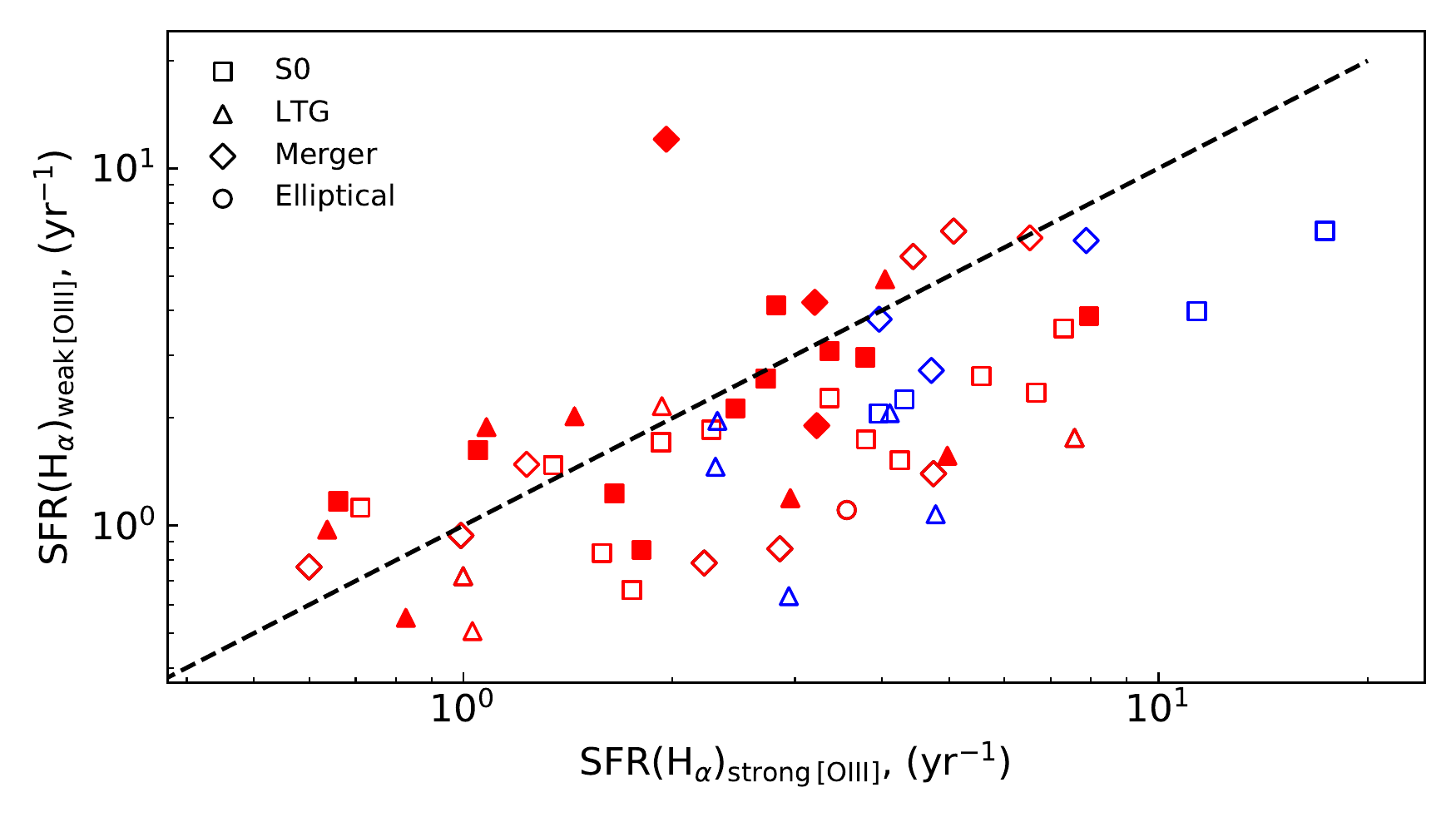}
  \caption{Star formation ratio calculated from the extinction corrected ${\rm H}_{\alpha}\lambda$6565 luminosity. On the x-axis (resp. y-axis), we display the SFR inferred from the strong {\rm [OIII]} (weak {\rm [OIII]}) component. We display the morphology (see Sect. \ref{ssec:morphology}) with different markers and indicate the BPT classification of the stronger {\rm [OIII]} line as colour-coded in Figure\,\ref{fig:first:bpt}. We highlight galaxies showing off-centred weak {\rm [OIII]} lines with empty markers. The dashed line corresponds to equal SFR in the two components. }
  \label{fig:sfr:halpha}
\end{figure}

We calculate the SFR from the extinction corrected ${\rm H}_{\alpha}\lambda$6565 luminosity as described in \citet{2002AJ....124.3135K}. We compute the extinction for each line component using the Balmer decrements \citep{2013ApJ...763..145D} assuming an intrinsic ratio ${\rm H}_{\alpha} / {\rm H}_{\beta} = 2.85$ \citep{1989agna.book.....O} and the Whitford reddening curve from \citet{1972ApJ...172..593M}. With the Balmer decrement, we estimate a mean E(B-V) of 0.6 for the two peaks. The measured mean [OIII]/$H_{\alpha}$ flux ratios of the weak and strong components are 0.2 and 1, while the differential dust attenuation between 5007A and 6565A is $0.7$. Hence, the relative reduction of one of the {\rm [OIII]} lines cannot be accounted for by extinction only. We subsequently compute the SFR following \citet{2002AJ....124.3135K}:
\begin{equation}
    {\rm SFR(H_{\alpha}) (M_{\odot} yr^ {-1}) = 7.9 \cdot 10^{-42} \,L_{\rm H_{\alpha}}}({\rm erg\,s^{-1}})
\end{equation}
In Figure\,\ref{fig:sfr:halpha}, we observe that the SFR associated to the strong {\rm [OIII]} component is larger than the weak {\rm [OIII]} one. This effect is even stronger for centred strong {\rm [OIII]} components (discussed in Sect. \ref{ssec:kinematics}) and especially for those which are also classified as composite (discussed in Sect. \ref{ssct:spectroscopic:selection}).

\section{Discussion}
\label{sect:disc}
The sample of emission-line galaxies discussed here exhibits two peaks lying in different regions of the BPT. We show that the gas velocities of the two peaks for only 22$\%$ of the galaxies are centred on the stellar velocity (see Fig. \ref{fig:offset:classification}).  All these galaxies exhibiting a symmetric kinematics are isolated but one (in a pair), and 69$\%$ of them are classified S0. This is not the expected proportion for the morphology of field disc galaxies \citep[e.g.][]{2009ApJ...702.1502V}. 78$\%$ of the galaxies have one gas peak associated to the stellar velocity, while the second one is offset. Only 33$\%$ of these galaxies with an offset kinematics are in pairs or small groups. All galaxies with the stronger {\rm [OIII]} component classified as composite (see Figure\,\ref{fig:first:bpt}) have a stellar velocity associated to this component (Figure\,\ref{fig:offset:classification}).

We do not find any significant extinction bias between the two peaks, and the optical snapshots do not reveals any asymmetric features. Similarly, one might think of off-centred circumnuclear discs like observed in NGC1068 \citep{2017A&A...608A..56G} or the off-centring of the AGN within the sphere of influence of the blackhole \citep{2019A&A...623A..79C}. However, it seems difficult to recover with a 3" fiber these relatively small scale features. Again non-resolved triaxial structures or bars in the central regions might also produce a large velocity gradient and possibly a double peak feature, but it would be difficult to account for the observed off-centring of one component. In addition, this cannot account for the unusual morphological types of this sample. 

On the one hand, half of the sample is composed of merger and late-type galaxies. As displayed in Figure\,\ref{fig:images}, it is difficult to disentangle a double nucleus if the galaxy is distant. It is thus possible that the number of mergers is underestimated. On the other hand, half of the sample is composed of S0 galaxies. While the number of S0 is estimated larger in clusters of galaxies (28$\%$ according to \citet{2009ApJ...702.1502V}), our sample is mainly composed of isolated galaxies or galaxies in small groups. 

Those galaxies are actively forming stars: they lie in the upper-mass range of the star forming main sequence (see Figure\,\ref{fig:ssfr:stellar:mass}). They are characterised by an enhanced star formation activity in their centre (see Figure\,\ref{fig:ssfr:ratio}). They also host an AGN. We find the SFR to be higher in the stronger {\rm [OIII]} component (Figure\,\ref{fig:sfr:halpha}). This effect is even more distinctive for strong {\rm [OIII]} components classified as composite which is counter intuitive since the weaker component is classified as SF.

Beside their morphological appearance, the S0 galaxies of this sample exhibit the same properties as the other galaxies of the sample. The origin of the S0 galaxies have been largely debated. While gas ejection by AGN has been proposed \citep[e.g.][]{2009ApJ...702.1502V}, it has also been shown that they could result from major/minor mergers \citep[e.g.][]{2018A&A...617A.113E}, which is supported here by the fact that some tidal features are observed around some of the S0 galaxies. These different points suggest that we might observe some mergers as well as ultimate phases of merging: galaxies with a double nucleus not resolved in the images or post-mergers (S0) with a central asymmetry. \citet{2018MNRAS.481.5580F} discuss that large-mass S0 galaxies might be formed by mergers, as studied by \citet{2018A&A...617A.113E}. Given the known co-evolution of SF and accretion of the black hole, the AGN activity can be concomitant with an enhancement of SFR (e.g. \citet{1994ApJ...425L..13M,2008A&ARv..15..189S}).

Last,  one can also discuss the possibility that the off-centred component is linked to an outflow of gas. Asymmetries in emission lines are known to be connected to gas outflow \citep{1981ApJ...247..403H,1985MNRAS.213....1W}. Such evidences are based on large field studies. \citet{2005ApJ...627..721G} and \citet{2016ApJ...817..108W} used double Gaussian emission line structures in the {\rm [OIII]} line only, while spectroscopic integral field unit studies show a difference in the velocity dispersion of the two lines \citep{2010ApJ...711..818S,2016ApJ...819..148K}, which is not the case here. These observations are consistent with measured outflows creating high offset velocity dispersion of around $1300 {\rm km s^{-1}}$ \citep{2013ApJ...768...75R}, even though smaller outflow velocities in the order of $100 {\rm km s^{-1}}$ have been observed in NGC 5929 \citep{2014ApJ...780L..24R} corresponding to double-peaks in the {\rm [OIII]} line. We cannot exclude that our observations correspond to an outflow of gas which might explain an offset component.

\section{Conclusion}
We presented a sample of 58 double-peaked galaxies, displaying a single strong peak in the ${\rm [OIII]}\lambda$5008 corresponding to one of the ${\rm H}_{\alpha}\lambda$6565 components. The two components are classified differently according to the BPT diagnostics, with one peak corresponding to an AGN or a composite region and the second one in the composite or SF region. In addition, we observe an off-centring of one of the components with respect to the stellar velocity in 78$\%$ of the galaxies. In addition, these massive galaxies ($\sim 10^{11}\,M_\odot$) are actively forming stars with a central enhancement, and 45$\%$ of them are S0 galaxies. The large majority (67$\%$) are isolated galaxies, while the others are hosted in small groups (with 2-5 galaxies), but one in a small cluster (of 27 galaxies). We can thus exclude this high fraction of S0 galaxies to be due to the environment. Given the galactic nuclei and star formation activities of these galaxies, we cannot exclude that we observe some gas outflows, which would require additional observations e.g. of molecular gas. In the meantime, it is probable that these kinematic signatures are linked to merging activity. 

\begin{acknowledgements}
We thank Fran\c coise Combes for interesting suggestions, and Gary Mamon for his support for this work. ALM has benefited from support from {\em Action F{\'e}d{\'e}ratrice "Cosmologie et Structuration de l'Univers"}. We thank the anonymous referee for constructive comments.\\

Funding for the SDSS and SDSS-II has been provided by the Alfred P. Sloan Foundation, the Participating Institutions, the National Science Foundation, the U.S. Department of Energy, the National Aeronautics and Space Administration, the Japanese Monbukagakusho, the Max Planck Society, and the Higher Education Funding Council for England. The SDSS Web Site is http://www.sdss.org/.\\

The Legacy Surveys (http://legacysurvey.org/) consist of three individual and complementary projects: the Dark Energy Camera Legacy Survey (DECaLS; NOAO Proposal ID \# 2014B-0404; PIs: David Schlegel and Arjun Dey), the Beijing-Arizona Sky Survey (BASS; NOAO Proposal ID \# 2015A-0801; PIs: Zhou Xu and Xiaohui Fan), and the Mayall z-band Legacy Survey (MzLS; NOAO Proposal ID \# 2016A-0453; PI: Arjun Dey). DECaLS, BASS and MzLS together include data obtained, respectively, at the Blanco telescope, Cerro Tololo Inter-American Observatory, National Optical Astronomy Observatory (NOAO); the Bok telescope, Steward Observatory, University of Arizona; and the Mayall telescope, Kitt Peak National Observatory, NOAO. The Legacy Surveys project is honoured to be permitted to conduct astronomical research on Iolkam Du`ag (Kitt Peak), a mountain with particular significance to the Tohono O'odham Nation.
\end{acknowledgements}

\appendix
\section{Additional figures}
\begin{wrapfigure}{r}{18cm}
\vspace{-0.5cm}
\includegraphics[width=17.5cm]{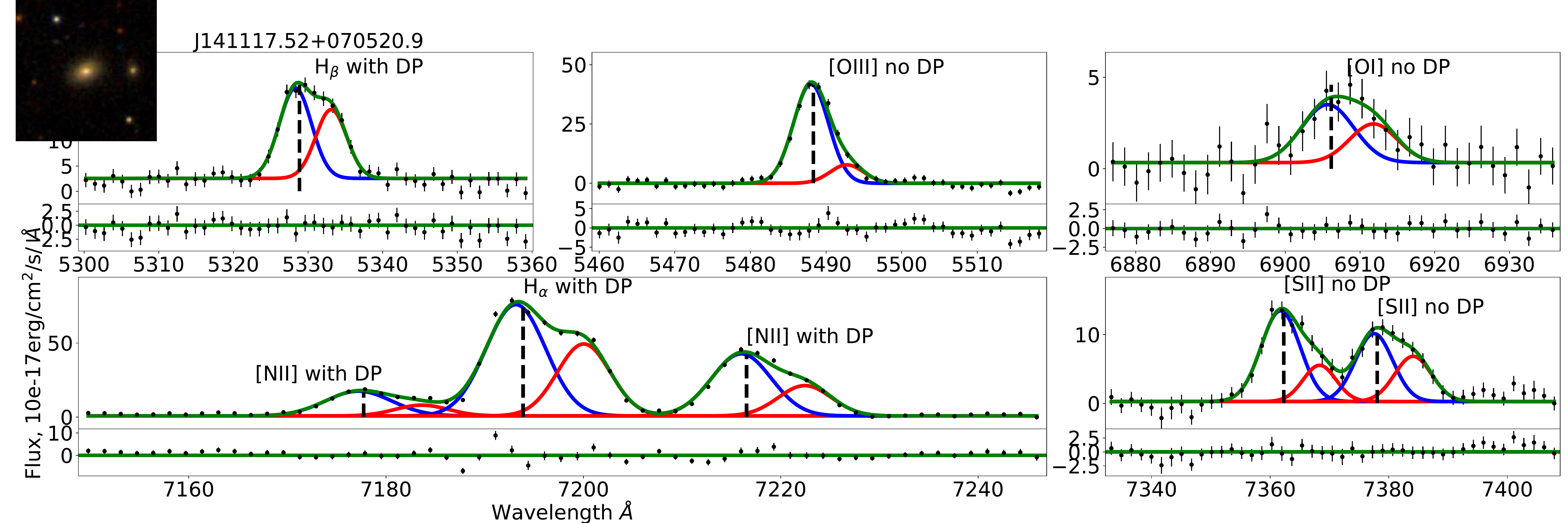}
\includegraphics[width=17.5cm]{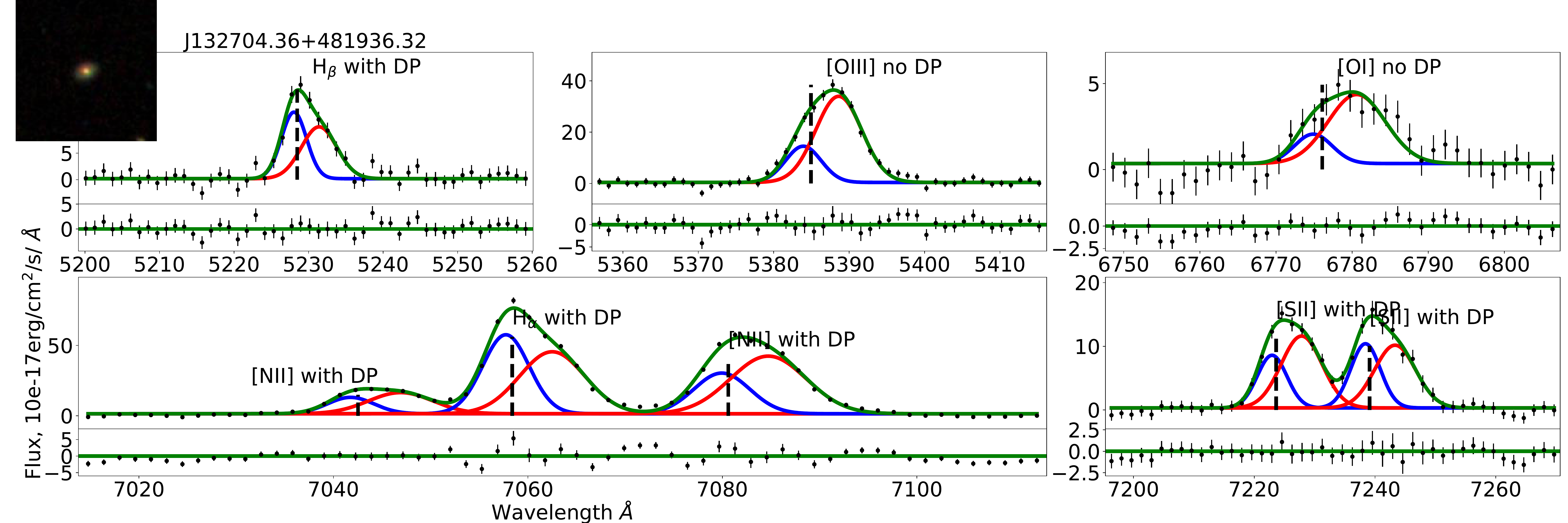}
\includegraphics[width=17.5cm]{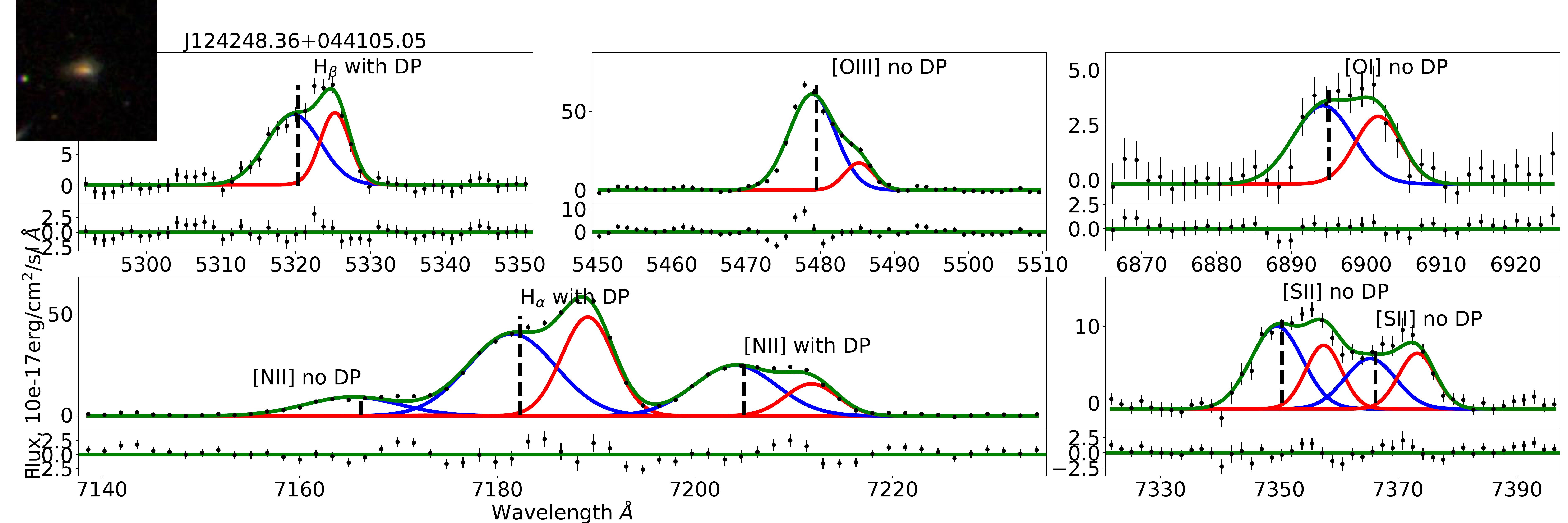}
\includegraphics[width=17.5cm]{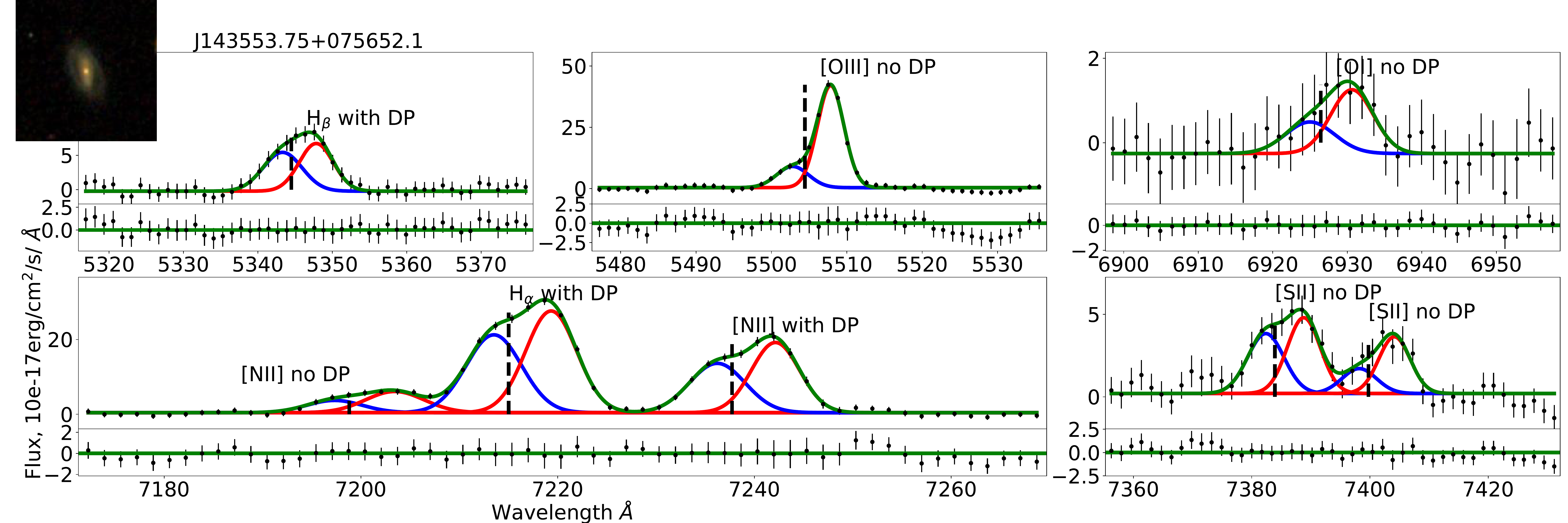}
\caption{Emission lines of other representative double-peak galaxies. The first two figures, on the top, are classified S0 while the last two figures correspond to LTG. Same colour coding as in Fig. \ref{fig:Spec}.}
    \label{fig:Spec2}
\end{wrapfigure}

\begin{table*}
\begin{minipage}[H][0.94\textheight][t]{\textwidth}
\renewcommand{\arraystretch}{0.96}
\begin{tabular}{l c c c c c c c c c c}   
\hline\hline         
ID & Designation & z & $\phi_{\rm s}^{[OIII]} $ & Morph & BPT & BPT& WHAN & WHAN & Kin & N$_{\rm g}$ \\ 
 & & & $ /\phi_{\rm w}^{[OIII]}$ &  & ${\rm [OIII]_{s}}$ & ${\rm [OIII]_{w}}$ & ${\rm [OIII]_{s}}$ & ${\rm [OIII]_{w}}$ & &  \\ 
(1) & (2) & (3) & (4) & (5) & (6) & (7) & (8) & (9) & (10) & (11)  \\ \hline
1 & J154403.67+044610.08 &0.042 & $ 4.8 \pm 0.7$ &\textit{Merger} & LINER & SF & sAGN & sAGN & off-w & 4 \\
2 & J075416.45+200136.32 &0.046 & $ 65.0 \pm 58.7$ &LTG & COMP & SF & sAGN & SF & off-w & 2 \\
3 & J091954.54+325559.79 &0.049 & $ 7.9 \pm 2.0$ &\textit{Merger} & COMP & SF & sAGN & SF & off-w & 0 \\
4 & J142859.54+605000.58 &0.052 & $ 13.4 \pm 7.6$ &S0 & LINER & COMP & wAGN & wAGN & cen & 1 \\
5 & J143555.01+103117.22 &0.055 & $ 4.6 \pm 1.2$ &LTG & AGN & COMP & wAGN & wAGN & off-s & 1 \\
6 & J143454.41-011618.06 &0.056 & $ 3.8 \pm 0.6$ &S0 & AGN & COMP & sAGN & wAGN & cen & 1 \\
7 & J084810.10+351534.05 &0.057 & $ 12.5 \pm 1.9$ &\textit{Merger} & AGN & COMP & sAGN & sAGN & off-w & 0 \\
8 & J120923.63+620955.96 &0.060 & $ 9.3 \pm 3.0$ &S0 & LINER & COMP & wAGN & wAGN & off-w & 1 \\
9 & J233141.99-103206.95 &0.061 & $ 3.8 \pm 1.1$ &S0 & LINER & COMP & wAGN & wAGN & cen & 1 \\
10 & J001119.84-093940.57 &0.062 & $ 79.5 \pm 30.0$ &LTG & AGN & SF & sAGN & RG & off-w & 1 \\
11 & J083515.21+511732.70 &0.067 & $ 5.0 \pm 1.6$ &\textit{Merger} & AGN & SF & sAGN & sAGN & off-w & 1 \\
12 & J153556.76-013749.17 &0.070 & $ 32.6 \pm 23.4$ &S0 & AGN & COMP & sAGN & wAGN & off-w & 1 \\
13 & J095600.70+130806.78 &0.070 & $ 15.9 \pm 6.7$ &LTG & COMP & SF & sAGN & SF & off-w & 2 \\
14 & J110233.35+224513.71 &0.074 & $ 16.6 \pm 10.0$ &LTG & LINER & COMP & wAGN & RG & off-w & 1 \\
15 & J164754.90+443345.05 &0.074 & $ 16.3 \pm 7.1$ &S0 & COMP & SF & sAGN & sAGN & off-w & 4 \\
16 & J142457.53+241517.82 &0.075 & $ 4.4 \pm 0.9$ &S0 & AGN & COMP & wAGN & wAGN & off-w & 2 \\
17 & J132704.36+481936.32 &0.076 & $ 3.1 \pm 0.6$ &S0 & AGN & COMP & sAGN & sAGN & off-w & 1 \\
18 & J142606.64+202831.56 &0.077 & $ 6.8 \pm 2.8$ &\textit{Merger} & COMP & SF & sAGN & SF & off-w & 4 \\
19 & J081204.73+171703.77 &0.081 & $ 4.0 \pm 0.9$ &\textit{S0} & AGN & COMP & sAGN & sAGN & cen & 1 \\
20 & J121446.40+013547.43 &0.083 & $ 8.1 \pm 3.1$ &S0 & AGN & COMP & wAGN & wAGN & off-w & 1 \\
21 & J015415.01+144716.78 &0.084 & $ 10.8 \pm 5.0$ &Ellip & AGN & COMP & sAGN & wAGN & off-w & 1 \\
22 & J165253.53+324440.29 &0.086 & $ 13.5 \pm 3.7$ &\textit{Merger} & AGN & COMP & sAGN & wAGN & off-w & 2 \\
23 & J090012.03+450514.82 &0.088 & $ 43.3 \pm 39.3$ &S0 & COMP & SF & sAGN & SF & off-w & 27 \\
24 & J073420.09+284038.96 &0.089 & $ 151.7 \pm 136.7$ &LTG & COMP & SF & sAGN & sAGN & off-w & 1 \\
25 & J093805.80+270456.72 &0.090 & $ 3.6 \pm 0.7$ &LTG & AGN & COMP & sAGN & sAGN & cen & 1 \\
26 & J131943.32+515255.83 &0.090 & $ 5.1 \pm 1.5$ &LTG & AGN & COMP & sAGN & sAGN & off-s & 2 \\
27 & J075932.35+193325.57 &0.093 & $ 15.0 \pm 7.3$ &S0 & AGN & COMP & wAGN & RG & cen & 1 \\
28 & J143035.27+443825.35 &0.093 & $ 9.8 \pm 3.7$ &LTG & COMP & SF & sAGN & wAGN & off-w & 1 \\
29 & J154124.59+271508.18 &0.094 & $ 9.8 \pm 3.8$ &S0 & LINER & COMP & sAGN & wAGN & off-w & 1 \\
30 & J124248.36+044105.05 &0.094 & $ 5.6 \pm 0.4$ &LTG & AGN & COMP & sAGN & SF & off-w & 1 \\
31 & J100759.81+630128.97 &0.095 & $ 3.1 \pm 0.6$ &LTG & AGN & COMP & wAGN & sAGN & off-s & 1 \\
32 & J141117.52+070520.89 &0.096 & $ 5.6 \pm 1.5$ &S0 & AGN & SF & sAGN & wAGN & off-w & 1 \\
33 & J143553.84+075652.13 &0.099 & $ 4.0 \pm 1.1$ &LTG & AGN & COMP & wAGN & wAGN & cen & 1 \\
34 & J113914.89-024107.87 &0.103 & $ 3.4 \pm 1.0$ &Merger & LINER & COMP & sAGN & wAGN & off-w & 1 \\
35 & J103554.15+110243.16 &0.105 & $ 8.5 \pm 1.3$ &\textit{Merger} & AGN & SF & sAGN & sAGN & off-w & 2 \\
36 & J144222.93+580434.42 &0.105 & $ 4.8 \pm 1.7$ &S0 & COMP & SF & sAGN & SF & off-w & 1 \\
37 & J075016.22+145816.70 &0.107 & $ 4.0 \pm 1.3$ &\textit{S0} & LINER & COMP & sAGN & sAGN & off-w & 0 \\
38 & J080738.22+504526.08 &0.120 & $ 13.9 \pm 6.3$ &\textit{Merger} & COMP & SF & sAGN & sAGN & off-w & 1 \\
39 & J154447.10+462817.72 &0.121 & $ 6.4 \pm 1.6$ &S0 & AGN & COMP & sAGN & wAGN & cen & 1 \\
40 & J142404.89+183606.76 &0.121 & $ 4.0 \pm 1.1$ &\textit{S0} & AGN & SF & wAGN & sAGN & cen & 1 \\
41 & J165742.02+380058.76 &0.122 & $ 5.3 \pm 1.0$ &Merger & AGN & SF & sAGN & sAGN & off-s & 1 \\
42 & J091431.22+551020.47 &0.123 & $ 4.7 \pm 1.2$ &S0 & AGN & COMP & wAGN & wAGN & cen & 1 \\
43 & J090211.82+083609.82 &0.128 & $ 3.5 \pm 1.1$ &LTG & AGN & COMP & sAGN & sAGN & off-s & 0 \\
44 & J164507.91+205759.50 &0.130 & $ 3.3 \pm 0.6$ &Merger & AGN & COMP & sAGN & sAGN & cen & 2 \\
45 & J164048.18+420042.75 &0.133 & $ 25.7 \pm 10.2$ &S0 & AGN & SF & sAGN & sAGN & off-w & 2 \\
46 & J090308.54+272639.04 &0.135 & $ 3.3 \pm 0.9$ &Merger & AGN & COMP & sAGN & sAGN & off-w & 1 \\
47 & J131044.32+245650.58 &0.136 & $ 74.0 \pm 29.6$ &LTG & COMP & SF & sAGN & wAGN & off-w & 1 \\
48 & J103510.83+420935.01 &0.137 & $ 10.4 \pm 4.0$ &\textit{S0} & COMP & SF & sAGN & SF & off-w & 1 \\
49 & J164155.60+252952.38 &0.137 & $ 8.6 \pm 3.6$ &LTG & AGN & COMP & sAGN & wAGN & off-w & 1 \\
50 & J105546.75+055409.21 &0.138 & $ 8.1 \pm 3.5$ &S0 & LINER & COMP & sAGN & wAGN & off-w & 1 \\
51 & J161050.02+324000.60 &0.140 & $ 14.9 \pm 7.7$ &S0 & AGN & COMP & sAGN & wAGN & off-w & 5 \\
52 & J104903.68+253323.97 &0.140 & $ 3.1 \pm 0.6$ &LTG & AGN & COMP & wAGN & sAGN & off-s & 1 \\
53 & J074715.76+370649.32 &0.140 & $ 4.0 \pm 0.9$ &Merger & AGN & COMP & sAGN & sAGN & cen & 1 \\
54 & J103945.99+173849.13 &0.143 & $ 6.9 \pm 2.4$ &S0 & AGN & COMP & sAGN & sAGN & off-w & 0 \\
55 & J171731.31+282424.67 &0.152 & $ 6.6 \pm 2.2$ &S0 & AGN & COMP & sAGN & sAGN & cen & 1 \\
56 & J150552.72+192909.16 &0.157 & $ 6.2 \pm 2.0$ &Merger & LINER & COMP & sAGN & sAGN & off-w & 1 \\
57 & J093610.57+303130.25 &0.162 & $ 17.7 \pm 11.8$ &Merger & AGN & SF & sAGN & sAGN & off-w & 1 \\
58 & J141426.56+643540.20 &0.166 & $ 5.3 \pm 1.8$ &S0 & AGN & SF & sAGN & sAGN & off-s & 0 \\

\hline 
\end{tabular}
\end{minipage}
\caption{Merger candidates sorted by redshift. (2) SDSS designation, (3) SDSS redshift in the third column, (4) ${\rm [OIII]}\lambda$5008 ratio between the strong {\rm [OIII]} and weak {\rm [OIII]} line component, (5) morphological classification, (6) and (7)  BPT diagram classification for strong (s) and weak (w) {\rm [OIII]} peak (Figure\,\ref{fig:first:bpt}), (8) and (9) WHAN diagram classification for each peak, (10)  kinematic classification of the gas emission line components with respect to the stellar velocity: 'off-s' (resp. 'off-w') off-centred strong (resp. weak) {\rm [OIII]} component and 'cen' for centred {\rm [OIII]} line components (see \ref{ssec:kinematics}), (11) number of identified galaxies in the corresponding group \citep{2007ApJ...671..153Y},  0 is provided when the galaxy has not been processed. }\label{table:properties}
\end{table*}

\clearpage
\newpage

\begin{figure}[!p]
  \centering 
 \includegraphics[width=0.48\textwidth]{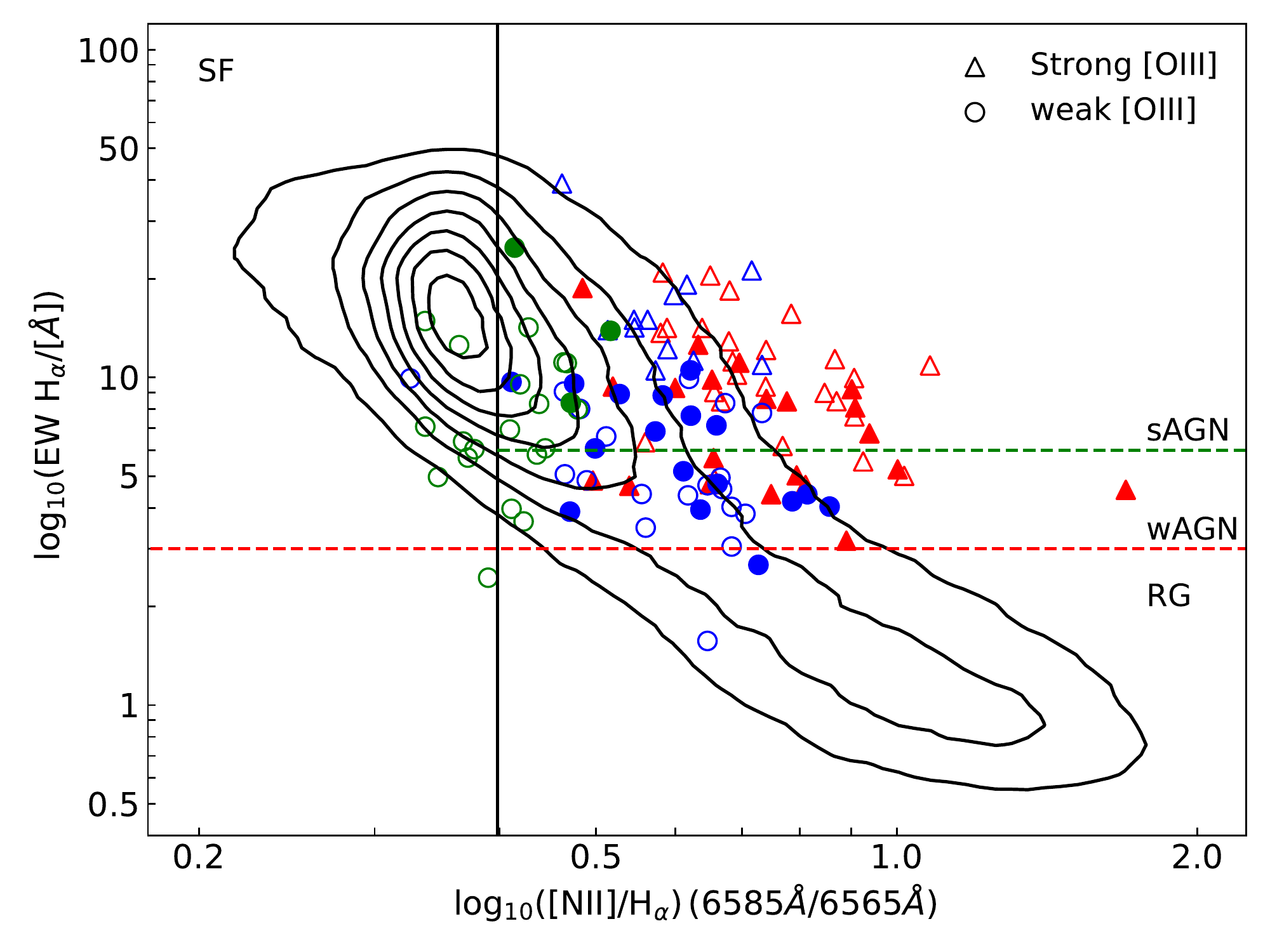}
  \caption{The diagnostic diagram introduced by \citet{2010MNRAS.403.1036C} as an alternative to the BPT classification (see Figure\,\ref{fig:first:bpt}). A topological separation classifies galaxies into star forming galaxies (SF), strong and weak AGNS (sAGN and wAGN) and retired galaxies (RG) \citep{2011MNRAS.413.1687C}. We compute the two double-peak components of the merger candidates individually: we display the component dominating the {\rm [OIII]}$\lambda$5008 line as triangles and the suppressed component as circles. We adapt the colour-coding according to the BPT classification in Figure\,\ref{fig:first:bpt}. We highlight galaxies with off-centred weak {\rm [OIII]} lines by empty markers (\ref{ssec:kinematics}) We observe that the majority of our classified components are situated in the AGN region. We recognise a shift between the two components which correlates with the {\rm [OIII]}$\lambda$5008 strength. The component dominating the {\rm [OIII]} is more likely classified as sAGN (45) and only some as wAGN (13). The suppressed {\rm [OIII]} component shows 26 sAGNs, 21 wAGNs 8 SF and 4 RG.}
  \label{fig:whan}
\end{figure}
\end{document}